\documentclass[prl,twocolumn,showpacs,amsmath,amssymb,superscriptaddress]{revtex4-1}

\usepackage{color}
\usepackage{amssymb}
\usepackage{amsthm}
\usepackage{graphicx}
\usepackage{epsfig}
\usepackage{subfigure}
\usepackage{amsmath,amssymb}
\usepackage{enumerate}
\usepackage[colorlinks=true,citecolor=blue,urlcolor=blue]{hyperref}
\usepackage{color}
\usepackage{bm}

\newcommand{\be}{\begin{equation}}
\newcommand{\ee}{\end{equation}}
\newcommand{\ba}{\begin{eqnarray}}
\newcommand{\ea}{\end{eqnarray}}
\newcommand{\ban}{\begin{eqnarray*}}
\newcommand{\ean}{\end{eqnarray*}}

\begin{document}

\title{\Large\textbf{Itinerant ferromagnetism of a dipolar Fermi gas with Raman-induced spin-orbit coupling}}
\author{Xue-Jing Feng}
\affiliation{Beijing National Laboratory for Condensed Matter Physics, Institute of Physics, Chinese Academy of Sciences, Beijing 100190, China}
\affiliation{School of Physics, Henan Normal University, Xinxiang 453000, China}
\author{Xing-Dong Zhao}
\affiliation{Beijing National Laboratory for Condensed Matter Physics, Institute of Physics, Chinese Academy of Sciences, Beijing 100190, China}
\affiliation{School of Physics, Henan Normal University, Xinxiang 453000, China}
\author{Lin Zhuang}
\affiliation{State Key Laboratory of Optoelectronic Materials and Technologies, School of Physics, Sun Yat-Sen University, Guangzhou 510275, China}
\author{Wu-Ming Liu}
\email{wmliu@iphy.ac.cn}
\affiliation{Beijing National Laboratory for Condensed Matter Physics, Institute of Physics, Chinese Academy of Sciences, Beijing 100190, China}
\affiliation{Songshan Lake Material Laboratory, Dongguan, Guangdong 523808, China}
\begin{abstract}
We elucidate the itinerant ferromagnetism of a dipolar Fermi gas with a Raman-induced spin-orbit coupling by investigating the exotic phase diagrams. It is revealed that the dipolar interaction along with spin-orbit coupling can corroborate the formation of ferromagnetism and the Raman coupling adversely eliminates the tendency to this ferromagnetism transition, which greatly transcends the general understanding of this subject with contact interaction only. We explore the ground states through the density and spin-flip distribution in momentum space, which exhibits novel degeneracy at strong Raman coupling indicated by a non-zero entropy at zero temperature. We calculate the transition temperatures well within the reach of an experimental system when altering the dipolar and spin-orbit coupling strength, which paves a way to the further experimental realization.
\end{abstract} 

\date{\today}
\maketitle
Itinerant ferromagnetism has been a subject of conspicuous interest in the history of physics. Early in last century when dealing with the itinerant electron gas within the Hartree-Fock approximation, Bloch pointed out that a ferromagnetic state can occur below a critical density at which the long-range Coulomb potential began to prevail over the kinetic energy. Thereafter Stoner studied the ferromagnetic properties in transition metals and gave a theoretical explanation \cite{stoner1933xxx,stoner1938collective} in which he replaced the Coulomb interaction with a screening repulsive contact potential. Subsequently in d-electron metals, the tight-binding model was commonly used including the single-band Hubbard model \cite{PhysRev.147.392,PhysRevLett.69.1608}. Unlike in a solid state system, a more rapidly developing quantum gas which is best known for its high tunability both in inner interactions and external magnetic or optical fields can provide a new experimental platform for and at the same time theoretically stimulate this intricate problem. \par
Experimental breakthrough came when the MIT group reported the investigation of the ferromagnetism transition in $^6$Li system \cite{jo2009itinerant}. Following that, however, was some theoretical dispute arguing that the experimental result was not convincing enough for no magnetic domains were captured and instead of the ferromagnetic state fermions might choose to be as interaction increased a short-range correlation state could also be a candidate to reduce the interaction energy \cite{PhysRevLett.10.159,PhysRevA.80.051605,PhysRevA.81.041602}. So further experimental explorations were carried out to verify the occurrence of the ferromagnetic state \cite{RN28,PhysRevLett.108.240404}. Meanwhile, many theoretical works made contributions to this subject. Beyond the mean-field approach, second-order perturbation calculation \cite{PhysRevLett.95.230403} was done, which obtained a critical phase transition point at $k_{F}a_{s}\approx 1.054$ where $k_{F}$ is the Fermi wave vector and $a_{s}$ is the $s$-wave scattering length. Other nonperturbative theoretical methods \cite{PhysRevA.85.043624,PhysRevA.93.063629,HE2014477} as well as the quantum Monte Carlo simulations \cite{PhysRevLett.103.207201,PhysRevLett.105.030405,Chang51} were also performed. In fact,  when we are studying the ferromagnetic instability of a ultra-cold Fermi gas, formation of molecules generated by three-body recombination \cite{zintchenko2016ferromagnetism,PhysRevA.83.043618} and the competing BCS pairing instability \cite{PhysRevLett.106.050402,PhysRevA.85.033628} shall be inevitably considered when this system undergoes a BEC-BCS crossover by tuning $a_{s}$ through Feshbach resonance. In another perspective, the occurrence of a ferromagnetic state could be seen as a spin-imbalanced circumstance \cite{PhysRevA.82.043626} in which a Fermi polaron was an interesting issue \cite{RN35,massignan2014polarons,PhysRevLett.110.230401}. Several works also found the mass imbalance in Fermi mixtures of which the usual two-component Fermi gas may be viewed as an equal-mass limit could stabilize the ferromagnetism \cite{RN35,PhysRevLett.110.165302,PhysRevA.83.053625}. Other focus on itinerant ferromagnetism were the explorations of dynamical properties in Fermi gas \cite{PhysRevLett.119.215303,PhysRevLett.104.220403,PhysRevLett.106.080402,PhysRevA.82.043603,PhysRevA.101.013618,sandri2011dynamical} as well as the non-equilibrium non-hermitian effect \cite{doi:10.7566/JPSJ.90.024004}. \par
 
However, most of the previous works in quantum gas were concentrating on an isotropic contact interaction as well as some finite-ranged and even higher partial-wave interactions \cite{2017JPhB...50a5302S,PhysRevA.85.033615,PhysRevB.87.184424,PhysRevA.98.023635,HE2014477}. Itinerant ferromagnetism induced by long-range and anisotropic dipole-dipole interaction (DDI) has been less investigated, by contrast, many unconventional quantum phases such as the supersolidity \cite{PhysRevB.89.174511}, charge and spin density waves \cite{PhysRevB.91.224504,PhysRevA.87.043604} were predicted in polar molecules $^{40}$K$^{87}$Rb \cite{K2008A,Bo2013Observation,PhysRevLett.108.080405,K2010Dipolar}, $^{23}$Na$^{40}$K \cite{PhysRevLett.109.085301} and magnetic dipolar $^{161}$Dy \cite{PhysRevLett.108.215301}. Apart from giving rise to the exhibition of exotic quantum phases, dipolar interaction also changed the shape of a spherical Fermi surface into a distorted one \cite{PhysRevA.77.061603,PhysRevA.81.033601,PhysRevLett.103.205301,20301,Feng_2020} and caused a structural second-order ferromagnetism transition \cite{PhysRevLett.103.205301,20301}. \par
In addition to the internal interaction, a common way of manipulating ultracold atoms is employing the external field to induce other interacting mechanics such as the spin-orbit coupling (SOC). As far as we know, SOC in condensed matter physics whose origination is the movement of an electron in an intrinsic electric field in a crystal is crucially responsible for numerous issues including topological insulators and Majorana fermions. While in cold atom physics, SOC arises from a synthetic gauge field created by the interaction between atoms and the Raman laser field \cite{RevModPhys.83.1523,Zhai2015Degenerate}.
Recently, one and two dimensional SOC have been successfully achieved in Bose and Fermi gas \cite{Victor2011,PhysRevLett.109.095301,science2016,PhysRevLett.117.235304,huang2016experimental}
 as well as in the dipolar fermion system \cite{PhysRevX.6.031022}. Theoretical explorations of two-dimensional Rashba SOC were reported in several papers \cite{PhysRevB.96.235425, PhysRevA.93.043602,VIVASC2020166113,PhysRevLett.108.080406,PhysRevLett.121.030404}, especially an interesting chiral ferromagnetism was demonstrated.

 \par
 \begin{figure}[t!]
 	\includegraphics[width=1\columnwidth]{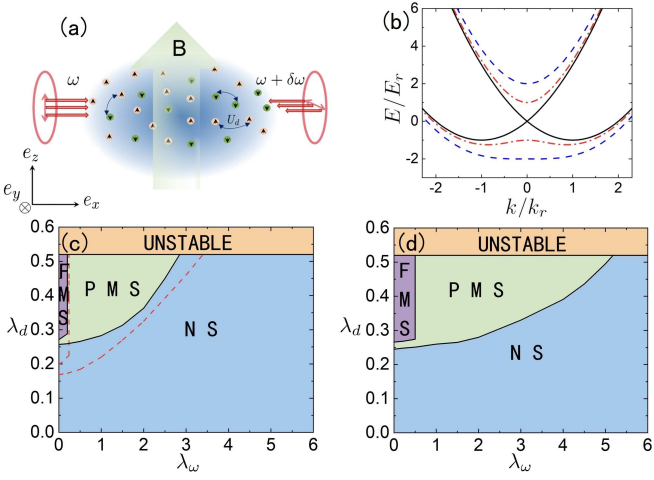}
 	\caption{(Color online) (a) Schematic representation of dipolar Fermi gas with a Raman-induced spin-orbit coupling. The magnetic field $\mathbf{B}$ is in $z$ direction and a pair of Raman lasers linearly polarized in $y$ and $z$ direction propagate through such an ensemble. (b) Dispersion spectrum for a Raman-induced spin-orbit coupled system without detuning. The black solid, red dash-dotted and blue dashed lines correspond to $\Omega=0$, $\Omega=2E_{r}$ and $\Omega=4E_{r}$. (c) Zero-temperature phase diagram as functions of Raman coupling strength $\lambda_{\omega}$ and dipolar interaction $\lambda_{d}$, with SOC parameter $\lambda_{\rm soc}=1$ and contact interaction $\lambda_{s}=0$. The phase diagram consists of four regions including the normal state (NS), partially magnetic state (PMS), fully magnetic state (FMS) and an unstable region. The dashed lines correspond to the phase boundaries shifted by a finite contact interaction $\lambda_{s}=0.5$. (d) is the same as (c), but for $\lambda_{\rm soc}=1.5$.}\label{Fig-1-torus}
 \end{figure}
 While in this work, a Raman-induced SOC is considered which is simply depicted in Fig.~1(a). As elucidated in Fig.~1(a), a magnetic filed in $z$ direction creates hyper-fine splitting for the spin-orbit coupling and a couple of $x$-direction Raman lasers that are polarized in $y$ and $z$ direction interact with cold atoms leading to an effective spin-orbit coupled Hamiltonian \cite{Zhai2015Degenerate}:
\begin{equation}
H_{\rm SOC}=\frac{\hbar^{2}k_{x}k_{0}}{m}\sigma_{z}+\frac{\delta}{2}\sigma_{z}+\frac{\Omega}{2}\sigma_{x},
\end{equation}
where $k_{0}$ is the wave vector of the laser, $\delta$ the Raman detuning parameter, $\Omega$ the Raman coupling, $k_{x}$ the $x$-direction momentum of the atom and $\sigma_{x}$, $\sigma_{z}$ are Pauli matrices. This effective Hamiltonian has a single particle dispersion relation depicted in Fig.~1(b) in which $E_{r}=\hbar^{2}k_{0}^{2}/(2m)$ is the recoil energy where $k_{0}$ is the wave vector of Raman lasers. As Raman coupling increases, the lowest double-well band evolves into a single-well shape, which, pointed out by Ref. \cite{Zheng_2013}, can explain the phase transitions among stripe phase, plane wave phase and non-magnetic phase in Bose gas. \par

\begin{figure}[t!]
	\includegraphics[width=1\columnwidth]{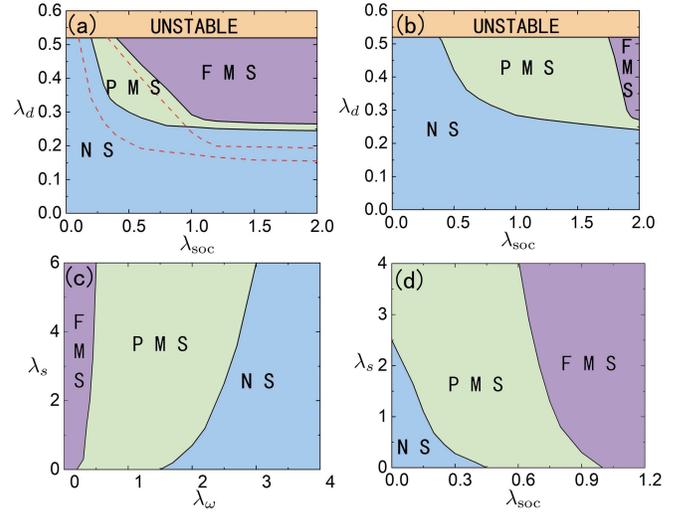}
	\caption{(Color online) (a) Zero-temperature phase diagram as functions of SOC strength $\lambda_{\rm soc}$ and dipolar interaction $\lambda_{d}$ with $\lambda_{\omega}=0.2$ and $\lambda_{s}=0$. The dashed lines correspond to the phase boundaries shifted by a finite contact interaction $\lambda_{s}=0.5$. (b) is the same as (a), but for $\lambda_{\omega}=1$. (c) Phase diagrams as functions of contact interaction $\lambda_{s}$ and Raman coupling strength $\lambda_{\omega}$ with $\lambda_{\rm soc}=1$. (d) Phase diagrams as functions of contact interaction $\lambda_{s}$ and SOC strength $\lambda_{\rm soc}$ with $\lambda_{\omega}=0.2$. Both (c) and (d) are for $\lambda_{d}=0.3$.}\label{Fig-symmetry-axis}
\end{figure}
\par
The Hamiltonian for the dipolar Fermi gas consists of three parts including kinetic part $H_{\mathbf{k}}$, SOC Hamiltonian $H_{\rm SOC}$ mentioned above and two-body interaction Hamiltonian $H_{I}$ which includes both dipolar interaction and the contact interaction.:
\begin{small}
\begin{equation}
H_{I}=\frac{1}{2}\int d^{3}\mathbf{x}d^{3}\mathbf{x^{'}}\psi_{\alpha}^ {\dagger}(\mathbf{x})\psi_{\beta}^{\dagger}
(\mathbf{x^{'}})U({\mathbf{x},\mathbf{x^{'}}})_{\alpha\alpha^{'},\beta\beta^{'}} \psi_{\beta^{'}}(\mathbf{x^{'}})\psi_{\alpha^{'}}(\mathbf{x}),
\end{equation}
\end{small}
\\ where $\psi_{\alpha}$ and $\psi_{\alpha}^{\dagger}$ are fermion annihilation and creation operator for the $\alpha$ component ($\alpha$ =1 and 2 represent spin-up and spin-down) and
\begin{small}
\begin{equation}\
U({\mathbf{x},\mathbf{x^{'}}})_{\alpha\alpha^{'},\beta\beta^{'}}=\frac{d^{2}}{r^{3}}\sigma_{\alpha \alpha^{'}}^{i}(\delta_{ij}-3\hat{\mathbf{r}}_{i}\hat{\mathbf{r}}_{j})\sigma_{\beta \beta^{'}}^{j}+g\delta_{\alpha\alpha^{'}}\delta_{\beta\beta^{'}}\delta(\mathbf{r}),
\end{equation}
\end{small}
\\ where $\hat{\mathbf{r}}\equiv (\mathbf{x}-\mathbf{x^{'}})/\mid \mathbf{x}-\mathbf{x^{'}}\mid$ and $d$, $g$ are the dipole moment of the fermions and the coupling strength of the contact interaction. \par
We apply a Hartree-Fock self-consistent method to study a dipolar Fermi gas with one-dimensional Raman-induced spin-orbit coupling (SOC). After a mean-field approximation and a canonical transformation we can obtain a set of self-consistent equations as displayed in the first section of supplemental materials. We introduce the dimensionless parameters including dipolar interaction
 parameter $\lambda_{d}=nd^{2}/\epsilon_{F}$, SOC parameter $\lambda_{\rm soc}=k_{0}/k_{F}$, contact interaction parameter $\lambda_{s}=gn/\epsilon_{F}$, Raman coupling parameter $\lambda_{\omega}=\Omega/\epsilon_{F}$  and temperature parameter $\lambda_{T}=k_{B}T/\epsilon_{F}$, where $\epsilon_{F}$, $k_{F}$, $k_{B}$ are Fermi energy, Fermi wave vector and Boltzmann constant, respectively. We denote $n_{\mathbf{k},\alpha}=\langle a_{\mathbf{k},\alpha}^{\dagger}a_{\mathbf{k},\alpha}\rangle$ as the particle density of spin-up and spin-down in momentum space and $t_{\mathbf{k}}=\langle  a_{\mathbf{k},\uparrow}^{\dagger}a_{\mathbf{k},\downarrow} \rangle$ as the spin-flip density.
\par
\begin{figure}[t!]
	\includegraphics[width=1\columnwidth]{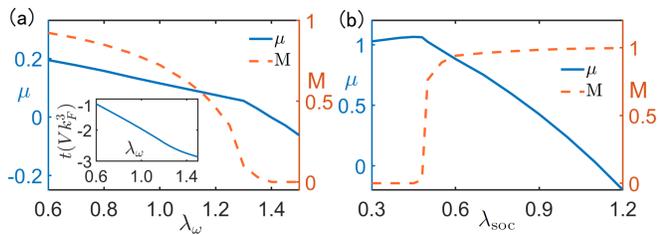}
	\caption{(Color online) (a) Zero-temperature chemical potential $\mu$ (blue solid line) and magnetisation M (orange dashed line) as functions of Raman coupling strength $\lambda_{\omega}$ with $\lambda_{\rm soc}$=1. The inset shows the total spin-flip $t=\sum_{\mathbf{k}}\langle a_{\mathbf{k},\uparrow}^{\dagger}a_{\mathbf{k},\downarrow}\rangle$ as a function of $\lambda_{\omega}$. (b) Zero-temperature chemical potential $\mu$ (blue solid line) and magnetisation M (orange dashed line) as functions of $\lambda_{\rm soc}$ with $\lambda_{\omega}$=0.2. Both (a) and (b) are obtained with $\lambda_{d}$=0.30 and $\lambda_{s}$=0.1.}\label{Fig-etg-Jpin}
\end{figure}
Our calculation indicates that ferromagnetism phase transition can occur under suitable parameters. We plot the phase diagrams as functions of $\lambda_{d}$, $\lambda_{\omega}$ and $\lambda_{\rm soc}$ shown in Fig.~1 and Fig.~2. An apparent conclusion can be drawn that Raman spin-flip effect can eliminate the tendency to ferromagnetic transition. The competition between dipolar interaction and Raman coupling might seem strange for spin-flip could intuitively imbalance the atoms of spin-up and spin-down thus favoring a ferromagnetic state. However, as we think further, the ground state should be a Hartree-Fock state with the following form:
\begin{equation}
\psi(\mathbf{r_{1},...,r_{N}})=\sum_{P}\frac{(-1)^{P}}{\sqrt{\rm N!}}\phi_{1}(\mathbf{r_{1}})\phi_{2}(\mathbf{r_{2}})...\phi_{\rm N}(\mathbf{r_{N}}),
\end{equation}
where $P$ is an arbitrary permutation. To put it more straightforward, we take $\rm N=2$ and the wave function with spin freedom becomes $\psi(\mathbf{r_{1},~r_{2},~\alpha,\beta})=\frac{1}{\sqrt{2}}[\phi_{1}(\mathbf{r_{1}},\alpha)\phi_{2}(\mathbf{r_{2}}, \beta)$-$\phi_{2}(\mathbf{r_{1}, \beta})\phi_{1}(\mathbf{r_{2}}, \alpha)]$. Considerring a symmetry-broken ferromagnetic state the wave function can be certainly written down as $\psi(\mathbf{r_{1},r_{2},\uparrow,\uparrow})$. If we regard the spin-flip term as an operator $\hat{F}$ satisfying $\hat{F}|\uparrow\rangle=|\downarrow\rangle, \hat{F}|\downarrow\rangle=|\uparrow\rangle$, then $\hat{F}$ has a zero expectation with $\psi(\mathbf{r_{1},r_{2},\uparrow,\uparrow})$. While for a normal state ($\rm S$=0), the wave function must be a combination of $\psi(\mathbf{r_{1},r_{2},\uparrow,\downarrow})$ and $\psi(\mathbf{r_{1},r_{2},\downarrow,\uparrow})$ and the expectation of $\hat{F}$ is not zero. The analysis above can be certainly generalized to a many-particle system. For a many-particle system, the expectation of $\hat{F}$ is zero even in a partially-ferromagnetic state and has a non-zero value only in a symmetric normal state. Thus a system with a spin-flip term favors a non-ferromagnetic phase. This effect can be also an analogy with the magnetic-nonmagnetic quantum phase transition as Raman coupling increases in a bosonic spin-orbit coupled system \cite{Zheng_2013}. \par
 \begin{figure}[t]
 	\includegraphics[width=1\columnwidth]{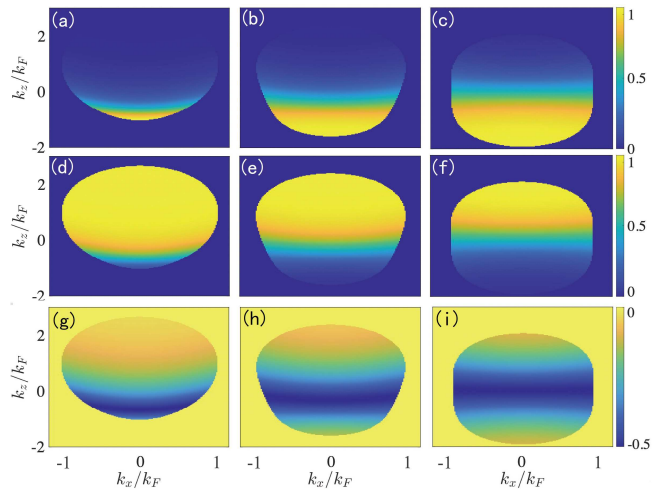}
 	\caption{(Color online) Zero-temperature density distribution $n_{\mathbf{k},\uparrow}$(a,b,c), $n_{\mathbf{k},\downarrow}$ (d,e,f) and spin-flip distribution $t_{\mathbf{k}}$ (g,h,i) with $\lambda_{\rm soc}$=1, $\lambda_{s}$=0.1, $\lambda_{d}$=0.30. (a), (d) and (g) are for $\lambda_{\omega}$=0.8; (b), (e) and (h) for $\lambda_{\omega}$=1.2; (c), (f) and (i) for $\lambda_{\omega}$=1.45. These figures from the left column to the right column shows a transition from a ferromagnetic state to a normal state.}\label{Fig-etg-Jpin}
 \end{figure}
It is also interesting from the phase diagrams of Fig.~2 that the 1-D SOC can enhance the ferromagnetism with a saturation, which can be seen from a rough calculation of the dipolar energy which takes the form of $(\rm \mathbf{d_{1}} \cdot \rm \mathbf{d_{2}})/r^{3}$$-\rm (\mathbf{d_{1}} \cdot \mathbf{r}) \rm (\mathbf{d_{2}} \cdot \mathbf{r})/r^{5}$ where $\mathbf{d_{1}}$, $\mathbf{d_{2}}$ and $\mathbf{r}$ are dipole moments and the separation of two dipoles. If we equally cast the dipoles of spin-up and spin-down into a spherical region, the interspecies DDI and intraspecies DDI cancel out. But if we separate two identical spherical balls each filled with dipoles of different spins, the interspecies DDI approaches to zero as the distance between two balls becomes large enough with the remaining intraspecies DDI a constant value. At a sufficiently large 1-D SOC, Fermi surfaces are well separated in momentum space, under which the difference of total DDI between a normal state and a ferromagnetic sate will saturate with the separation of Fermi surfaces.\par
To know how contact interaction influences the phase diagram, we plot the phase diagrams with finite contact interaction which are displayed in Fig~1 and Fig~2. So we can conclude that the unusual contact interaction can promote the formation of a magnetic state, which is vastly studied in the previous papers \cite{PhysRevLett.95.230403,PhysRevA.93.063629}. So is the contact interaction the only crucial necessity for the ferromagnetism transition? The answer is no. As pointed out in this work, the dipolar interaction as well as the SOC plays a substituted role in determining the spontaneous polarization.\par
The detail of the ferromagnetic transition shall be analyzed by plotting the chemical potential and magnetization as displayed in Fig.~3. The derivations of chemical potential at transition points behave discontinuously manifesting a first-order ferromagnetic phase transition. The order parameters $n_{\mathbf{k},\uparrow}$ and $n_{\mathbf{k},\downarrow}$ are depicted in Fig.~4 which have a rotational symmetry. Quite contrary to an ideal spherically Fermi surface, the distribution of particles in momentum space shows a distorted shape because of the presence of anisotropic dipolar interaction. On the other hand, the shapes of Fermi surfaces are also influenced by the Raman coupling strength whose detail can be referred to Fig.~S1 of the supplemental materials. Interestingly, Raman coupling leads to a non-zero spin-flip $t_{\mathbf{k}}$ which has a non-uniform distribution in momentum space shown in Fig.~4. The total spin-flip $t=\sum_{\mathbf{k}}t_{\mathbf{k}}$ is a negative value and declines monotonously as $\lambda_{\omega}$ increases which is displayed in Fig.~3. We can regard this spin-flip distribution as a symmetry ``gate" through which particles of spin-up can accumulate and particles of spin-down can escape. As $\lambda_{\omega}$ increases further, this ``gate" becomes more widespread which makes the zero-temperature ground state a pseudo-symmetric one different from the general Pauli paramagnetic state. Here we have to specify the pseudo-symmetric normal state in our phase diagrams as a combination of a true normal state ($\rm S$=0) and an $x$-direction polarized state. A true normal state minimizes the kinetic energy and an $x$-direction polarized state minimizes the Raman coupling energy. Thus as $\lambda_{\omega}$ increases, the ground state should be a combination of an actually normal state and $x$-direction polarized state which minimizes the total energy. \par
 As the dipolar interaction  increases, there displays a dynamical unstable properties. In this unstable region, compressibility $K^{-1}=n(\partial P/\partial n)$ becomes negative where pressure $P=-(\partial E/\partial V)_{N}$. What has to be pointed out is that the boundary line of dynamical unstable region doesn't rely on $\lambda_{\omega}$ or $\lambda_{\rm soc}$ and is hammered at $\lambda_{d}\approx 0.52$ \cite{PhysRevLett.103.205301}, which can be inferred from the following facts. When $\lambda_{\omega}$ is small enough, the state near the unstable boundary is a fully magnetic state and $t_{\mathbf{k}}$ equals to zero thus leading to none contribution to the total energy. When $\lambda_{\omega}$ is large enough, the state near the unstable boundary is a fully $x$-direction polarized state and $t=\sum_{\mathbf{k}}t_{\mathbf{k}}$ is a constant. The energy of Raman coupling part takes the form of $\Omega Vk_{F}^{3}t$ whose second derivative to $n$ is zero thus also making no contribution to compressibility. As for the intermediate region, Raman coupling term equals to an $x$-direction exerted magnetic field and doesn't influence the intrinsic unstable properties as we have argued in my previous paper \cite{20301} that a momentum-dependent magnetic field in $z$ direction doesn't change the unstable region. 
\par
\begin{figure}[t]
	\includegraphics[width=3.6in]{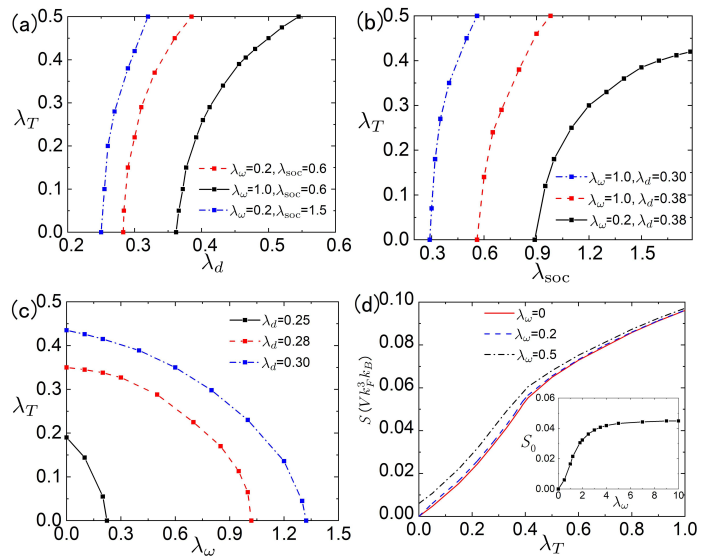}
	\caption{(Color online) Ferromagnetic transition temperature as functions of dipolar interaction $\lambda_{d}$ (a), SOC parameter $\lambda_{\rm soc}$ (b) and Raman coupling strength $\lambda_{\omega}$ (c). (a) and (b) are for $\lambda_{s}$=0 and (c) for $\lambda_{s}$=0.1 and $\lambda_{\rm soc}$=1. (d) Entropy as functions of temperature $\lambda_{T}$ with $\lambda_{d}$=0.30, $\lambda_{\rm soc}$=1 and $\lambda_{s}$=0.1. The inset is the zero-temperature entropy varying with $\lambda_{\omega}$, which saturates to a certain value at large $\lambda_{\omega}$.}\label{Fig5}
\end{figure}

In Fig.~5, we plot the ferromagnetic transition temperature as functions of $\lambda_{d}$, $\lambda_{\rm soc}$ and $\lambda_{\omega}$. Transition temperature increases with $\lambda_{d}$ and $\lambda_{\rm soc}$ and declines with $\lambda_{\omega}$. Finally, it is also of great interest to know how entropy behaves at finite temperature which takes the form of $S=-k_{B}\sum_{\mathbf{k}} \bigg[f(\mathbf{k})\ln f(\mathbf{k})
+(1-f(\mathbf{k}))\ln(1-f(\mathbf{k}))\bigg]$. \par
As displayed in Fig.~5(d), entropy increases as temperature increases, which accords with our general knowledge. Interestingly however, the entropy is not always zero as temperature approaches zero and its value has an increasing dependence on $\lambda_{\omega}$ and saturates at a certain value at large $\lambda_{\omega}$. This zero-temperature entropy's attaining to zero is valid according to the third law of thermodynamics. While in quantum statistics, zero-temperature entropy is usually related to the degeneracy of ground states. As we mentioned above, Raman coupling results in a spin-flip distribution in momentum space which becomes nearly uniform as zero-temperature entropy attains its saturation value. 
\par
In most of the previous experiments, two-component fermions were usually a mixture of ultracold $^{6}$Li atoms \cite{jo2009itinerant,RN28,PhysRevLett.108.240404} in which the system could be cooled down to about 0.1$T_{F}$ to 1$T_{F}$. By tuning the effective scattering length $a_{s}$ through Feshbach resonance, a strong repulsive branch could be reached in which a Stoner-type itinerant ferromagnetism could be possibly verified. In a recent Raman spin-orbit coupled dipolar $^{161}$Dy system \cite{PhysRevX.6.031022}, the Zeeman sublevels of $|\downarrow \rangle \equiv |F=21/2,m_{F}=-21/2 \rangle$ and $|\uparrow \rangle \equiv |F=21/2,m_{F}=-19/2 \rangle$ are coupled by two Raman lasers with wavelength $\lambda=741 \rm nm$. The parameters of $\lambda_{\omega}$ and $\lambda_{\rm soc}$ are about 1 and 0.4 with the peak density of $10^{14}$ cm$^{-3}$. The dipolar interaction parameter $\lambda_{d}$ is about 0.02 and the temperature $\lambda_{T}$ ranges from 0.1 to 0.4. To observe this ferromagnetic transition demonstrated in our work, apart from manipulating the Raman lasers, we can manage to increase the effective dipolar interaction \cite{li2021tuning}. To observe a spin polarization experimentally, monitoring the suppression of collision could be an adopted way as collisions would be forbidden in a fully ferromagnetic state \cite{jo2009itinerant}. Otherwise a probing of the spin-dipole dynamics can also demonstrate the spin susceptibility \cite{RN28}. The predicted deformation of the Fermi surfaces can be also easily explored by a free expansion method \cite{aikawa2014observation}.
\par
In summary, we have investigated the itinerant ferromagnetic phase transition in a Raman-induced spin-orbit coupled dipolar Fermi gas, which is mainly dominated by the long-range dipole-dipole interaction. The presence of Raman-induced spin-orbit coupling makes great contributions to the formation the itinerant ferromagnetism and provides us a feasible tool to manipulate the system. The long-range dipole-dipole interaction and the spin-orbit coupling also bring us new physical mechanisms, for instance, the deformations of the two Fermi-surfaces which can be different in the ferromagnetic phase and become the same in the paramagnetic state. The high possibility of experimental observation comes from the fact that our theoretical models can be related to the dipolar $^{161}$Dy system.
\par
This work was supported by the National Key R$\&$D Program of China under grants No. 2021YFA1400900, 2021YFA0718300, 2021YFA1400243, NSFC under grants Nos.  61835013.
\bibliography{paper3}



\end{document}